\documentclass[11pt]{article}
\usepackage{fullpage}
\usepackage{xspace}
\usepackage{amsmath}
\usepackage{latexsym}
\usepackage{amsthm}
\usepackage{amssymb}
\usepackage{url}
\usepackage{graphicx}
\usepackage{booktabs}
\usepackage{newalg}
\usepackage{epsfig}
\usepackage{layout}

\begin{document}
\newtheorem%
     {theorem}{Theorem}[section]
\newtheorem%
     {corollary}[theorem]{Corollary}
\newtheorem%
     {proposition}[theorem]{Proposition} 
\newtheorem%
     {lemma}[theorem]{Lemma} 

\newtheorem%
     {exampleAux}[theorem]{Example}

\newenvironment%
     {example}{\begin{exampleAux}\rm}{\qed\end{exampleAux}}

\newenvironment%
     {examplenq}{\begin{exampleAux}\rm}{\end{exampleAux}}

\newtheorem%
     {examplesAux}[theorem]{Examples} 
\newenvironment%
     {examples}{\begin{examplesAux}}{\end{examplesAux}}

\newtheorem{definition}[theorem]{Definition}

\newtheorem%
     {constructionAux}[theorem]{Construction} 
\newenvironment%
     {construction}{\begin{constructionAux}\rm}{\end{constructionAux}}

\newcounter{inv}

\newenvironment%
     {invariant}{\begin{list}{INV--\arabic{inv}:}{\usecounter{inv}\setlength{\labelwidth}{2.8cm}\setlength{\leftmargin}{1.9cm}}}{\end{list}}

\def\proof{{\sl Proof.\ \ }}
\def\proofsketch{{\sl Proof (Sketch).\ \ }}

\def\qed{\hfill{\boxit{}}
  \ifdim\lastskip<\medskipamount \removelastskip\penalty55\medskip\fi}
\long\def\boxit#1{\vbox{\hrule\hbox{\vrule\kern3pt
                  \vbox{\kern3pt#1\kern3pt}\kern3pt\vrule}\hrule}}

\newenvironment%
     {genass}%
        {\medbreak\noindent{\bf General Assumption.\enspace}\it}%
        {\ifdim\lastskip<\medskipamount \removelastskip\penalty55\medskip\fi}


\newcommand {\boxfigure}[1]%
   {\framebox[\textwidth]{%
    \parbox {0.99\textwidth}
                {{#1}\vspace {0cm}\hfill}}}

\newcommand {\boxfigureone}[1]%
   {\framebox[\textwidth]{%
    \parbox {0.90\textwidth}
                {{#1}\vspace {0cm}\hfill}}}

\newcommand{\gefig}[3]
        {\begin{figure}[htb] 
        \begin{center}%
        \mbox{}
        {\psfig{figure=#1,width=0.60\textwidth}}
        \mbox{}
        \end{center}
        \caption{{#2}\label{#3}} 
        \end{figure}}

\newcommand{\hangif}[1]{\raisebox{-1ex}{\hspace{2em}
                                        \makebox[1.0em][l]{if}
                                        \parbox[t]{320pt}{#1}}}


\def\A{{\cal A}} \def\B{{\cal B}} \def\C{{\cal C}} \def\D{{\cal D}}
\def\E{{\cal E}} \def\F{{\cal F}} \def\G{{\cal G}} \def\H{{\cal H}}
\def\I{{\cal I}} \def\J{{\cal J}} \def\K{{\cal K}} \def\L{{\cal L}}
\def\M{{\cal M}} \def\N{{\cal N}} \def\O{{\cal O}} \def\P{{\cal P}}
\def\Q{{\cal Q}} \def\R{{\cal R}} \def\S{{\cal S}} \def\T{{\cal T}}
\def\U{{\cal U}} \def\V{{\cal V}} \def\W{{\cal W}} \def\X{{\cal X}}
\def\Y{{\cal Y}} \def\Z{{\cal Z}}

\newcommand{\dd}[2]{#1_1,\ldots,#1_{#2}}      

\newcommand{\set}[1]{\{#1\}}
\newcommand{\eset}{\emptyset}
\newcommand\bigset[1]{ \Bigl\{ #1 \Bigr\} }   
\newcommand\bigmid{\ \Big|\ }

\newcommand{\incl}{\subseteq}           
\newcommand{\incls}{\supseteq}          

\newcommand{\col}{\colon}

\newcommand{\NP}{{\rm NP}}              
\newcommand{\GI}{{\rm GI}}              
\newcommand{\PIPEETWO}{\Pi^{{\rm P}}_2} 
\newcommand{\SIGPEETWO}{\Sigma^{{\rm P}}_2}     
\newcommand{\PSPACE}{{\rm PSPACE}}      
\newcommand{\EXPTIME}{{\rm EXPTIME}}    

\newcommand{\angles}[1]{\langle#1\rangle}       


\newcommand{\OnlyIf}{\lq\lq$\Rightarrow$\rq\rq\ \ }   
\newcommand{\If}{\lq\lq$\Leftarrow$\rq\rq\ \ }        

\newcommand{\quotes}[1]{\lq\lq#1\rq\rq}         
\newcommand{\wrt}{w.r.t.}                       
\newcommand{\WLOG}{w.l.o.g.}                    
\newcommand{\ie}{i.e.}                          
\newcommand{\eg}{e.g.}                          
\renewcommand{\hom}{homo\-mor\-phism}

\newcommand{\eat}[1]{}

\newcommand{\comment}[1]{\noindent{\it COMMENT:} {\it #1}}

\newcommand{\paramP}{\langle\P\rangle}
\newcommand{\gcon}{\subset_{\text{\sc is}}}
\newcommand{\gconeq}{\subseteq_{\text{\sc is}}}
\newcommand{\PIO}{{\rm PIO}\xspace}
\newcommand{\PINC}{{\rm PINC}\xspace}
\newcommand{\PTIME}{{\rm PTIME}\xspace}

\newcommand{\undpath}[3]{#1\stackrel{#2}{\longleftrightarrow} #3}
\newcommand{\dpath}[3]{#1\xrightarrow{#2} #3}
\newcommand{\notV}{\overline V}
\newcommand{\graphPreceq}{\subseteq}
\newcommand{\graphPrec}{\subset}
\newcommand{\funTime}[2]{t_{{}_\P}(#1, #2)}
\newcommand{\funExtend}[2]{e_{{}_\P}(#1, #2)}
\newcommand{\funSize}[2]{s_{{}_\P}(#1, #2)}

\newcommand{\funExtendVert}[2]{e'_{{}_\P}(#1, #2)}
\newcommand{\funSizeVert}[2]{s'_{{}_\P}(#1, #2)}
\renewcommand{\setminus}{-}
\newcommand{\cupSub}{\Cup}
\newcommand{\alter}[1]{{\it alter}(#1)}
\newcommand{\vrank}[2]{{\it rank}(#1,#2)}
\newcommand{\gtn}[2]{{>_{{}_{#2}}}(#1)}
\newcommand{\ltn}[2]{{<_{{}_{#2}}}(#1)}
\newcommand{\gless}[1]{<_{{}_{#1}}}
\newcommand{\Peel}[3]{{\sl Peel}(#1,#2,#3)}

\newcommand{\VCS}{\mbox{\sc VCS}\xspace}
\newcommand{\IPVCS}{\mbox{\sc IncPolyVCS}\xspace}
\newcommand{\CHP}{\mbox{\sc GenHered}\xspace}
\newcommand{\IPCCHP}{\mbox{\sc IncPolyCCHP}\xspace}
\newcommand{\IPCRHP}{\mbox{\sc IncPolyCRHP}\xspace}
\newcommand{\DAGCCHP}{\mbox{\sc DAG\_CCHP}\xspace}
\newcommand{\CRHP}{\mbox{\sc CRHP}\xspace}
\newcommand{\DCRHP}{\mbox{\sc DagCRHP}\xspace}
\newcommand{\TCCHP}{\mbox{\sc TreeCCHP}\xspace}
\newcommand{\OCP}{\mbox{\sc OrderedCP}\xspace}
\newcommand{\CCHP}{\mbox{\sc CCHP}\xspace}

\newcommand{\bigcard}[1]{\bigl|#1\bigr|}
\newcommand{\extend}[3]{{\it Ext}_{#3}(#1,#2)}
\newcommand{\restrRank}[2]{[#1]_{#2}}
\newcommand{\rank}[1]{{\it rank}(#1,\P)}
\newcommand{\Trav}[1]{{\it Tr}(#1)}
\newcommand{\delV}[2]{#1_{\hat #2}}
\newcommand{\maxSolRank}[3]{{\it MSol}^{#3}(#1,#2)}
\newcommand{\maxSol}[2]{#1^m(#2)}
\newcommand{\maxSolVert}[3]{#1^m(#2;\, #3)}
\newcommand{\sol}[2]{#1(#2)}
\newcommand{\solRank}[3]{{\it Sol}^{#3}(#1, #2)}
\newcommand{\inducedSub}[1]{2^{#1}}
\newcommand{\layer}[2]{{\it Layer}(#1,#2)}
\newcommand{\emptyGraph}{O_0}
\newcommand{\prop}[1]{{\sf ``#1''}}
\newcommand{\plusNode}{\cup} 
\newcommand{\plusNodeG}[1]{\cup_{{}_{#1}}} 

\title{Generating All Maximal Induced Subgraphs for Hereditary, 
  Connected-Hereditary and Rooted-Hereditary Properties} 
\author{Sara Cohen\thanks{The  Selim and Rachel Benin School of
  Computer Science and Engineering, The Hebrew University of
  Jerusalem, Givat Ram Campus, Jerusalem 91904, Israel. Email: {\tt
  \{sarina,sagiv\}@cs.huji.ac.il}} \and Yehoshua Sagiv$^*$
}
\date{}
\maketitle

\thispagestyle{empty}
\begin{abstract}

The problem of computing all maximal induced subgraphs of a graph $G$
that have a graph property $\P$, also called the {\em maximal $\P$-subgraphs
  problem}, is considered. This problem is studied for hereditary,
connected-hereditary and rooted-hereditary graph properties.
The maximal $\P$-subgraphs problem is
reduced to restricted versions of this problem by providing algorithms
that solve the general problem, assuming that an algorithm for a
restricted version is given. The complexity of the algorithms are
analyzed in terms of total polynomial time, incremental polynomial
time and the complexity class P-enumerable. The general results
presented allow simple proofs that the maximal $\P$-subgraphs problem
can be solved efficiently (in terms of the input and output) for many
different properties. 
\eat{
, e.g.,  bipartite
graphs. In databases, various problems, such as  full
disjunctions, can be computed as maximal solutions to some
connected-hereditary properties.  
For hereditary, connected-hereditary and rooted-hereditary
properties, it is shown that under some 
 general conditions, all maximal
solutions can be generated in polynomial time under input-output complexity.
If some additional conditions are satisfied, then  all maximal
solutions can be generated in incremental 
polynomial time, and $k$ maximal solutions can be generated in
polynomial time. 
The above general results include  and improve upon previous results in
databases, in particular, and in graph theory, in general. For
example,  it is shown that full disjunctions can
be computed in incremental polynomial time, and that maximal bipartite
subgraphs can be generated in polynomial time under input-output complexity. 
}


%

\eat{
This paper presents algorithms that generate all maximal solutions for
hereditary, connected-hereditary and rooted-hereditary graph
properties. For {\em polynomially-extendable\/} and {\em polynomially
  vertex-extendible properties\/}, our algorithms run in polynomial
time under input-output complexity.  It turns out that usually
properties that are {\em polynomially solvable}, i.e., properties for
which it is possible to generate all maximal solutions in polynomial 
 time under input-output complexity, are in one of the two classes
 that we have identified. Properties
  for which maximal solutions can be generated in incremental
  polynomial time and for which $K$ maximal solutions can be generated
  in polynomial time, for a given constant $K$, are also identified. 

 The problem of finding ``maximal query answers'' over
 incomplete databases has been studied in the past in various
 different settings. It turns out that many of methods for computing
 maximal query answers  can be modeled  abstractly as hereditary,
 connected-hereditary or rooted-hereditary  properties. 
Hence, our algorithms yield 
  immediately as special cases several complexity results that were
  proven independently in the past for the problem of finding
  ``maximal query answers.''  In addition, we improve on previous
  results in this area by providing 
  incremental polynomial algorithms that solve these problems. Our
  algorithms also provide polynomial solutions to open problems in
  graph theory, such
  as that of generating all maximal bipartite subgraphs of a given graph.

}
\eat{Our goal is to generate all maximal solutions for a variety of
problems in polynomially time under input-output complexity. Instead
of presenting an algorithms for each problem, we consider the very
general classes of  hereditary,
 connected-hereditary and rooted-hereditary graph properties, and
 present algorithms for each class of problems. 
}


\eat{We consider the problem of generating maximal solutions for 
hereditary, connected-hereditary and rooted-hereditary properties over
graphs. 

Contributions: 
* We recognized the fact that many problems for which we want to find
maximal answers can be modeled abstractly as hereditary,
connected-hereditary or rooted-hereditary properties. 
* Formulated abstract algorithms 
answers can be 
}
\end{abstract}

\section{Introduction} \label{sec:intro}
\eat{
A graph property $\P$ is a set of graphs. For
example, the set of cliques, line graphs and 3-colorable graphs are
graph properties. Hereditary graph properties are graph properties
that are closed with respect to induced subgraphs. Such graph properties
include many common types of graphs, such as those mentioned
above. Connected-hereditary properties are graph properties that are
closed with respect to  connected induced subgraphs and are also common. 
Rooted-hereditary properties are graph properties that are
closed with respect to rooted induced subgraphs and appear in
database theory in the context of semistructured data. 
}

Hereditary and connected-hereditary graph properties include many common
types of graphs such as  cliques, bipartite  graphs
and trees. Such properties appear in many contexts, and thus, they have
been widely studied, e.g.,~\cite{Lewis:Yannakakis:JCSS,Yannakakis:Node:Edge:Deletion,Balogh:Speed-Hereditary,Sos:Saks-Diversity}.%
\eat{The node-deletion problem for hereditary properties has been
proven  to be NP-complete~\cite{Lewis:Yannakakis:JCSS}. 
This is also true of the node-deletion problem for
connected-hereditary
properties~\cite{Yannakakis:Node:Edge:Deletion}. The speed of 
growth~\cite{Balogh:Speed-Hereditary} 
and the diversity~\cite{Sos:Saks-Diversity} of hereditary properties
have also been analysed.
}
 This paper focuses on the  {\em maximal $\P$-subgraphs problem}: Given a
 graph property $\P$  and an arbitrary graph  $G$, find all maximal induced
 subgraphs of $G$ that have the property $\P$. 
 We consider properties 
 that are hereditary, connected hereditary or rooted hereditary (a
 variant of connected hereditary). 

Since the output for the maximal $\P$-subgraphs problem may be large,
our complexity analysis takes into consideration both the size of the
input and the size of the output. Specifically, we consider the
complexity measures {\em total polynomial
 time\/}~\cite{Johnson-Yannakakis:Generating-Maximal-Independents-Sets} and
 {\em incremental polynomial
 time\/}~\cite{Johnson-Yannakakis:Generating-Maximal-Independents-Sets},
 and the complexity class {\em
 P-enumerable\/}~\cite{Valiant:Complexity:Computing:Permanent}.

The maximal $\P$-subgraphs problem has been studied for many
properties $\P$. For example, it has been shown that this problem is
both P-enumerable and solvable in incremental polynomial time 
for the properties  \prop{is an independent set} and  \prop{is
 a
 clique}~\cite{Johnson-Yannakakis:Generating-Maximal-Independents-Sets,Tsukiyama-Maximal:Independant:Sets,Akkoyunlu-Maximal:Cliques}.   
If subgraphs (and not only induced subgraphs) are allowed, then the
maximal $\P$-subgraphs problem is P-enumerable for the properties 
\prop{is a spanning
  tree}~\cite{Read:Tarjan-Listing:Cycles:Paths:Spanning:Trees,Kappor:Ramesh-Generating:Spanning:Trees}, 
\prop{is an elementary
  cycle}~\cite{Read:Tarjan-Listing:Cycles:Paths:Spanning:Trees,Johnson-Elementary:Ciruit} 
and \prop{is an elementary
  path}~\cite{Read:Tarjan-Listing:Cycles:Paths:Spanning:Trees,Babic:Graovac-Enumeration:Walks}, 
among others.

This paper differs from previous work in that we do not consider
specific properties, but instead, deal with the problem for a general
$\P$. Our strategy is to reduce the maximal $\P$-subgraphs problem to
restricted versions of the  maximal $\P$-subgraphs problem that are
often easier to solve. Our reductions are by means of algorithms
that solve the maximal $\P$-subgraphs problem, given a solution to a 
restricted version of this problem. 

\eat{We believe
that our approach is promising, since it involves defining algorithms
for a class of properties, instead of  for 
specific properties.
Our formulation allows our algorithms to be immediately
applied to new problems.
These general results include and
improve upon previous results in graph theory. For example, it is
shown that the maximal \prop{bipartite}-subgraph problem is solvable
in total polynomial time, which improves
upon~\cite{Number:Bipartite:Graphs}.   
}

Using our algorithms, we can show that the maximal $\P$-subgraphs
problem can be solved in total-polynomial time, incremental-polynomial
time or is  P-enumerable if certain conditions hold on the runtime
of a restricted version of this problem. Hence, our approach is easily
shown to include and improve upon previous results in graph
theory. For example, it is 
shown that the maximal \prop{bipartite}-subgraph problem is solvable
in total polynomial time, which improves
upon~\cite{Number:Bipartite:Graphs}.   

The maximal $\P$-subgraphs problem also has immediate practical
applications in the database field. Interestingly, it turns out that
many well-known semantics for answering queries in the presence of
incomplete information can be modeled as hereditary,
connected-hereditary or rooted-hereditary graph properties. Hence, the
results in this paper imply and improve upon the complexity results
in~\cite{Kanza:Et:Al-Incomplete:Answers:Over:SSDs-PODS,Kanza:Sagiv:Full:Disjunctions,Cohen:Generating:Relation} 
and imply the complexity result
in~\cite{Yannakakis-Algorithms:Acyclic:Database:Schemas-VLDB}. In
fact, modeling semantics as graph properties allows previously
presented semantics to be extended without affecting their
complexity. See~\cite{ICDT:Submission} for more details.

\eat{
This paper differs from previous work in that we present a general
algorithm that allows for simple proofs that the Maximal $\P$-Subgraphs
Problem is P-enumerable, if certain conditions hold. If some
additional conditions are satisfied, then the  Maximal $\P$-Subgraphs
Problem can be solved in incremental polynomial time, and $k$ maximal
$\P$-subgraphs can be generated in polynomial time, for a given
constant $k$. The above general results include  and improve upon
previous results in graph theory. For example,  it is shown that the
Maximal Bipartite-Subgraph problem is P-enumerable.

Databases often contain incomplete information. This is especially
common when the database is formed by integrating information from
heterogeneous sources or in semistructured databases. In such
situations, there may not be enough 
information to completely answer a query. Hence, it is of interest to
find maximal answers for a given query. Different semantics for
computing maximal answers have been proposed
for a variety of settings
(e.g., full disjunctions~\cite{Galindo-Legaria:Full:Disjunctions},
weak and {\sc
  or}-semantics~\cite{Kanza:Et:Al-Incomplete:Answers:Over:SSDs-PODS},
maximally interconnected nodes~\cite{Cohen:Generating:Relation}).
These semantics differ  significantly
one from another. Nonetheless,  it has been
shown that each of these semantics is {\em polynomially-solvable\/},
i.e., it is possible to generate all maximal solutions under each of
these semantics in polynomial time under input-output complexity. 

Obviously, when defining a semantics for maximal answers 
it  is extremely desirable that the semantics be
polynomially-solvable. However,  a full understanding of what
causes a semantics to be polynomially-solvable is lacking. Hence, it is
difficult to discern whether some newly defined semantics is in fact
polynomially-solvable. Actually, determining whether a semantics is
polynomially-solvable may be quite intricate. For example, although full
disjunctions were introduced in
1994~\cite{Galindo-Legaria:Full:Disjunctions}, only recently was it
determined that full disjunctions 
are polynomially-solvable~\cite{Kanza:Sagiv:Full:Disjunctions}. (A
special case was 
shown to be polynomially-solvable in
1996~\cite{Rajaraman:Ullman-Computing:Full:Disjunctions-PODS}.)
This makes it difficult to define a new polynomially-solvable 
semantics for maximal answers for a language such as XQuery.

One of the contributions of this paper is in determining that the
above semantics 
can be couched as computing maximal solutions for a connected-hereditary or
rooted-hereditary graph property. It turns out that this is the underlying
reason why the semantics are polynomially-solvable. 
Therefore, in this paper we study the problem of computing maximal
solutions for 
hereditary, connected-hereditary and rooted-hereditary properties. We
are interested, in particular, in polynomially-solvable properties. 
In contrast to previous work in this area, we do not
provide algorithms for specific properties. Instead, we identify
general conditions for which it is possible to generate all maximal
solutions in polynomial time under input-output complexity.

Our general results include  and improve upon previous results in
databases, in particular, and in graph theory, in general. For
example,  it is shown that full disjunctions can
be computed in incremental polynomial time, and that maximal bipartite
subgraphs can be generated in polynomial time under input-output
complexity. In  Section~\ref{sec:conclusion} we present detailed
examples of problems that can be modeled as hereditary,
connected-hereditary or rooted-hereditary properties. We also show
how our results improve upon previous results for these problems. 
}
\eat{
This paper is organized as follows.  
Section~\ref{sec:definitions}
presents necessary definitions. 
Some background on computing maximal solutions is presented in
Section~\ref{sec:characteristics}. 
In Section~\ref{sec:hereditary}
an algorithm that generates maximal solutions for hereditary
properties is presented. 
In Section~\ref{sec:connected-hereditary} an
algorithm that generates maximal solutions for connected-hereditary
and rooted-hereditary properties is presented. An incremental polynomial
algorithm, that generates solutions for hereditary,
connected-hereditary and rooted-hereditary properties, is presented in
Section~\ref{sec:ip:connected-hereditary}.
Section~\ref{sec:conclusion} concludes.
}
\eat{
We
identify classes of properties for which all maximal solutions can be
generated in polynomial time under input-output complexity, and
present algorithms that solve this problems. 

For hereditary, connected-hereditary and rooted-hereditary
properties, this paper gives
 general conditions guaranteeing that all maximal
solutions can be generated in polynomial time under input-output complexity.
If some additional conditions are satisfied, then  all maximal
solutions can be generated in incremental 
polynomial time, and $k$ maximal solutions can be generated in
polynomial time.

Therefore, we study the general problem of computing maximal solutions...
Results.
Results contain and improve.
Interestingly, our results also solve open problems in graph theory.

It is often

Hereditary, connected-hereditary and rooted-hereditary graph
properties include many common graph properties. 
Informally, a property is hereditary if the fact that it holds on a
graph $G$, implies that it holds on all induced subgraphs of
$G$. Similarly, a property is connected-hereditary or
rooted-hereditary if the fact that it holds on $G$ implies that it
holds on all connected or rooted induced subgraphs of $G$.  
In databases, the problem of computing maximal answers (rather than
complete answers) of a query over incomplete information can be
couched as computing maximal solutions for a connected-hereditary or
rooted-hereditary property. 

 The problem of finding ``maximal query answers'' (rather than complete
 answers) over
 incomplete information has been studied in the past in various
 different settings. This problem 
 can often be couched as computing maximal solutions for a
 connected-hereditary or rooted-hereditary property.

It turns out that many of methods for computing
 maximal query answers can be modeled  abstractly as hereditary,
 connected-hereditary or rooted-hereditary  properties. 
For example,
In databases, various problems, such as  full
disjunctions, can be computed as maximal solutions to some
connected-hereditary property.

}

\eat{
It is often the case that a database contains only incomplete
information. 
In such situations, one may be interested in
computing {\em maximal solutions\/} for a given query, instead of
computing complete solutions for the query. For relational
databases, maximal solutions are often computed by means of full
disjunctions~\cite{}.
In~\cite{Kanza:Et:Al-Incomplete:Answers:Over:SSDs-PODS}, the problem
of computing maximal 
solutions for graph queries over semi-structured databases was
studied. \cite{Cohen:Generating:Relation} focused on the problem of
  generating maximal relations from XML documents. The number of
  maximal solutions for a given query may be exponential in the size
  of the input
  (i.e., the query and the database). 
Hence, one is often
  interested in the input-output complexity of such problems, i.e., in
  the complexity as a function of the input and the output. 
In each
  of~\cite{,Kanza:Et:Al-Incomplete:Answers:Over:SSDs-PODS,Cohen:Generating:Relation},
  algorithms that compute maximal solutions in polynomial time under
  input-output complexity, were presented.

We have shown
that these problems can be modeled abstractly as hereditary,
connected-hereditary or rooted-hereditary properties over graphs. We then
consider the general problem of computing maximal solutions for
hereditary, connected-hereditary and rooted-hereditary 
properties. We identify the classes of properties called {\em
  polynomially-extendable properties\/} and {\em polynomially
  vertex-extendible properties}. It is often easy to verify whether a
given property is in one of these classes. For both
polynomially-extendable and polynomially vertex-extendible properties we
present algorithms that compute maximal solutions in polynomial time
under input-output complexity. It turns out that the problems
mentioned in the previous paragraph 
are in these classes of properties. Hence, as special cases, our
algorithms show that the problems above can be solved in polynomial
time under input-output complexity. Our algorithms can also be used to 
}
\eat{
Often we are interested in finding maximal answers for a given
problem. For example, one 

We consider the problem of generating all maximal solutions for
hereditary, connected-hereditary and rooted-hereditary graph
properties. Informally, a 
property is hereditary if the fact that it holds on a graph $G$,
implies that it
holds on all induced subgraphs of $G$. Similarly, a property is
connected-hereditary if the fact that it holds on  $G$ implies that it
holds on all connected induced subgraphs of $G$.  Finally, a property is
rooted-hereditary if the fact that it holds on  $G$ implies that it
holds on all rooted induced subgraphs of $G$. 

Hereditary properties, and their variants, are common in graph
theory. For example, the 
property \prop{is a clique}
is hereditary. Many algorithms have been developed
that generate all maximal solutions for specific hereditary
properties. In our paper, we consider properties with abstract
characteristics, such as polynomially extendible properties and
polynomially vertex-extendible properties. We present
algorithms that generate all maximal solutions for hereditary,
connected-hereditary and rooted-hereditary properties with these
characteristics. Our 
algorithms are polynomial under input-output complexity. We also
present an algorithm that runs in polynomial incremental  time. 

The algorithms presented improve on some known results for generating
maximal solutions for specific properties. In addition, we believe
that our approach is promising, since it involves defining algorithms
for a class of properties, instead of  for 
specific properties.
Our formulation allows our algorithms to be immediately
applied to new problems.

This paper is organized as follows. Section~\ref{sec:definitions}
presents necessary definitions. In Section~\ref{sec:examples}, a
variety of examples of hereditary, connected-hereditary and
rooted-hereditary properties are
discussed. Some background on the problem is presented in
Section~\ref{sec:characteristics}. These
characteristics are used in Section~\ref{sec:hereditary} to formulate
an algorithm that generates maximal solutions for hereditary
properties. The characteristics are also used in
Section~\ref{sec:connected-hereditary} in an  
algorithm that generates maximal solutions for connected-hereditary
and rooted-hereditary properties. A polynomial incremental algorithm,
that generates solutions for hereditary, connected-hereditary and
rooted-hereditary properties, is presented in
Section~\ref{sec:ip:connected-hereditary}. Some special cases where
maximal solutions for connected-hereditary and rooted-hereditary
properties can be generated in linear time in the size of the output
are presented in Section~\ref{sec:linear:time}.
}

\section{Graphs and Graph Properties}
\label{sec:definitions}


\paragraph{Graphs and Induced Subgraphs.}
A {\em graph} $G=(V,E,r)$
consists of {\em (1)\/} a finite set 
of {\em vertices\/} $V$, {\em (2)\/} a set of {\em edges\/}
$E\subseteq V\times V$ {\em and 
  (3)\/} a {\em root\/} 
$r$ such that $r\in V\cup\set{\bot}$. 
We say that $G$ is {\em rooted\/} if {\em (1)\/} $r\neq \bot$  {\em
  and (2)\/} every vertex in $G$ is reachable via a directed path from
 $r$. \label{rooted} 
We say that $G$ is {\em connected\/} if its underlying undirected
graph is connected. 
Observe  that every rooted graph is
  connected. However, a connected graph need not be rooted. 
We use $V(G)$ to denote the set of vertices of $G$.

  A graph $H $ is an {\em induced subgraph\/} of a graph
  $G$, written  $H\gconeq G$, if {\em (1)\/} $H$ is
  derived from $G$ by deleting some of the vertices of $G$ (and the edges
  incident on these vertices) {\em and (2)\/} $H$ has the same root as
  $G$, if the root of $G$ is among the vertices of $H$, and has $\bot$
  as its root otherwise.
We write $H\gcon G$ if $H\gconeq G$
  and $H$ is not equal to $G$. 

We use $G[\set{\dd v n}]$ to denote the induced subgraph
of $G$ that contains exactly the vertices $\dd v n$. If $H$ and $H'$
are induced subgraphs of $G$ and $v$ is a vertex in $G$, we use $G[H]$, $G[H,v]$ and $G[H,H']$ as
shorthand notations for $G[V(H)]$, $G[V(H)\cup \set{v}]$ and $G[V(H)\cup
V(H')]$, respectively.

\paragraph{Graph Properties.}
A {\em graph property} $\P$ is a nonempty and possibly 
infinite set of graphs. 
For example, \prop{is a clique} is a graph 
property that contains all graphs that are
cliques. In this paper, we only consider properties $\P$ such that it
is  possible to verify whether a graph $G$ is in $\P$ in polynomial
time. Hence, we assume that there is a polynomial procedure  
  {\sc Sat}$\paramP$ that receives a graph $G$ as input,  
and returns true if $G\in\P$ and false otherwise. Observe that  the
notation $\paramP$  denotes an algorithm that is
parameterized by the graph property $\P$, i.e., that differs for each
value of $\P$. 

We consider several special types of graph properties.
A graph property $\P$ is
{\em hereditary\/} if $\P$ is {\em closed\/} with respect to induced
subgraphs, i.e., whenever $G\in\P$, 
every induced subgraph of $G$ is also in $\P$. 
A graph property $\P$ is {\em connected hereditary\/} if 
{\em (1)\/} all the graphs in $\P$ are connected {\em and (2)\/}
$\P$ is closed with respect to  connected induced subgraphs.
%
A graph property $\P$ is {\em rooted-hereditary\/} if 
{\em (1)\/} $\P$ only holds on rooted graphs {\em and (2)\/}
$\P$ is closed with respect to rooted induced subgraphs.\footnote{This type of
properties is useful when considering database problems related to
semistructured data, since semistructured data are usually represented
as rooted graphs.}
It is rather unusual for a rooted-hereditary
property to also be hereditary or connected-hereditary. 
Actually, one can show that if $\P$ is rooted-hereditary and $\P$ is
also hereditary or connected-hereditary, then $\P$ contains only
graphs with at most one vertex.
This gives an
additional motivation to considering 
  rooted-hereditary properties, since they
generally differ from connected-hereditary and hereditary properties.

Many graph properties are
 hereditary~\cite{garey-johnson:intractability}, e.g.,  
 \prop{is a clique} 
 and \prop{is a
  forest.} Note that 
 \prop{is a clique} is also 
connected hereditary. However,  \prop{is a clique} is not
rooted hereditary, since it contains graphs that do not have roots.
Some properties are connected hereditary, but
not hereditary or rooted hereditary, such as  \prop{is a
  tree,} which contains a graph $G$ if
 the underlying undirected graph of $G$ is a tree. Note that $G$ is
 not necessarily rooted. Hence, \prop{is a tree} is not rooted
 hereditary.
The property \prop{is a rooted clique} is
rooted hereditary.

\paragraph{The Maximal $\P$-Subgraphs Problem.}
  Let $G$ be a graph and $\P$ be a property. (The graph $G$ is not
  necessarily in $\P$.) We say that $H$ is a {\em 
  $\P$-subgraph\/} of $G$ if $H\gconeq G$ and
  $H\in \P$. The set of $\P$-subgraphs of a graph $G$ is denoted
  $\sol \P G$. 

We say that $H$ is  a {\em maximal $\P$-subgraph\/} of $G$ if
  $H$ is a $\P$-subgraph of $G$ and there is no $\P$-subgraph $H'$ of
  $G$, such that  $H\gcon H'$. We use $\maxSol \P G$ to denote the set
  of maximal $\P$-subgraphs of $G$. 
The {\em maximal $\P$-subgraphs
  problem\/} is: Given a graph $G$, find the set $\maxSol \P G$.

\newcommand{\SIZE}[1]{|#1|}
\newcommand{\OUT}{\text{\sc out}}

\section{Complexity Classes and Measures} 
This paper explores the problem of computing $\maxSol \P G$, for
a hereditary, connected-hereditary or rooted-hereditary property $\P$
and an arbitrary graph $G$. 
The maximal $\P$-subgraphs problem cannot be solved in polynomial
time, in the general case. This follows from the fact that sometimes
the size of $\maxSol \P G$ is exponential in the size of
$G$ (see~\cite{Balogh:Speed-Hereditary} for details). Hence, exponential
time may be needed just to print the output.%
\eat{
Consider the following naive procedure that computes  $\maxSol \P
G$:
\begin{itemize}
\item {\bf Step 1:} Compute the set $\sol \P G$.
\item {\bf Step 2:} Remove from $\sol \P G$ any graph that is not maximal.
\end{itemize}
Obviously, this procedure is too expensive to be of interest. For
example, if $\P$ is hereditary and $G\in \sol \P G$, then 
 a set of size $2^{|V(G)|}$ will be created in the first
step. During the second step, all graphs except $G$ will be
removed. Thus, the procedure will have run in exponential time, even
though its output is quite small. To summarize, the main fault of the naive
procedure lays in the fact that is always runs in exponential time,
regardless of the size of the output. 
}
In this section, we discuss two complexity measures 
 that are of interest when the output of a problem may be
large: {\em total
  polynomial
 time\/}~\cite{Johnson-Yannakakis:Generating-Maximal-Independents-Sets}
 and  {\em incremental polynomial
 time\/}~\cite{Johnson-Yannakakis:Generating-Maximal-Independents-Sets}.
 We also consider the complexity class {\em
  P-enumerable\/}~\cite{Valiant:Complexity:Computing:Permanent}.  

A problem can be solved in {\em total polynomial time}, or {PIO}
for short, if the time 
required to list all its solutions is bounded by a polynomial in 
$n$ (the size of the input) and $K$ (the number of solutions in the
output).\footnote{This complexity measure is similar to polynomial
  time input-output complexity, which is commonly considered in
  database theory,
  e.g.~\cite{Yannakakis-Algorithms:Acyclic:Database:Schemas-VLDB}.} 
 For the maximal $\P$-subgraphs problem, $n$ is the
number of vertices in $G$ and $K$ is the number of graphs in $\maxSol
\P G$. 

The complexity class {\em P-enumerable\/} is more restrictive than 
the measure of total polynomial time. Formally, a problem is
P-enumerable if  the time required to list all its solutions is
bounded by $K$ times a polynomial in $n$. Note that P-enumerable
differs from total polynomial time in that the factor of the output in
the runtime must be linear. Since the size of the output may be
exponential in the size of the input, 
the factor of output size in the
total runtime is highly influential.

Another complexity measure that is of interest when dealing with
problems that may have large output (such as the
maximal $\P$-subgraphs problem) is {\em incremental polynomial
  time}.
Formally, a problem is  solvable in incremental polynomial time, or
PINC for short,  if,
for all $k$, the $k$-th solution of
the output  can be returned in polynomial time in $n$ (the
input) and $k$. 
Incremental polynomial time is of importance when the user would like
to optimize  evaluation time for retrieval of the first $k$
maximal induced subgraphs, as opposed to optimizing for overall time. 
This is particularly useful in a scenario where the user reads the
answers as they are delivered, or is only interested in looking at a
small portion of the total result. If a problem is solvable in
total polynomial time, but not in incremental polynomial time, 
 the user may have to wait exponential time until
the entire output is created, before viewing a single maximal
$\P$-subgraph. 

Observe that every problem that is P-enumerable is also
solvable in total polynomial time. Similarly, every problem that is
solvable in incremental polynomial time is also solvable in total
polynomial time. It is not known whether every problem that is
solvable in incremental 
polynomial time is also P-enumerable, and vice-versa. 
The maximal $\P$-subgraphs problem has been studied for many
properties $\P$. See Section~\ref{sec:intro} for several examples and
see~\cite{Combinatorial:Enumeration:Website} for a
listing of algorithms for combinatorial enumeration problems.  
\eat{For example, it has been shown that this problem is
both P-enumerable and solvable in incremental polynomial time 
for the properties  \prop{is an independent set} and  \prop{is
 a
 clique}~\cite{Johnson-Yannakakis:Generating-Maximal-Independents-Sets,Tsukiyama-Maximal:Independant:Sets,Akkoyunlu-Maximal:Cliques}.   
If subgraphs (and not only induced subgraphs) are allowed, then the
maximal $\P$-subgraphs problem is P-enumerable for the properties 
\prop{is a spanning
  tree}~\cite{Read:Tarjan-Listing:Cycles:Paths:Spanning:Trees,Kappor:Ramesh-Generating:Spanning:Trees,Gabow:Myers-Finding:Spanning:Trees}, 
\prop{is an elementary
  cycle}~\cite{Read:Tarjan-Listing:Cycles:Paths:Spanning:Trees,Johnson-Elementary:Ciruit} 
and \prop{is an elementary
  path}~\cite{Read:Tarjan-Listing:Cycles:Paths:Spanning:Trees,Babic:Graovac-Enumeration:Walks}, 
among others. See~\cite{Combinatorial:Enumeration:Website} for a
listing of algorithms for combinatorial enumeration problems.  
}

\newcommand{\PCBIP}{\P_{\text{\sc cbip}}}
\newcommand{\PBIP}{\P_{\text{\sc bip}}}

\section{\mathversion{bold}Restricting the Maximal $\P$-Subgraphs Problem}
Let $\P$ be a graph property. Suppose that we want to show that the
maximal $\P$-subgraphs problem is in PIO. To do this   
we must devise an algorithm that, when given any graph $G$,
produces $\maxSol \P G$ in polynomial time in the
input (i.e., $G$) and the output (i.e., $\maxSol \P G$). 
For many properties $\P$, it is difficult to find such an algorithm,
since an arbitrary graph $G$ must be dealt with. 
%
Our task of finding
an appropriate algorithm is even more
difficult if we actually want to show that the maximal $\P$-subgraphs
problem is P-enumerable or is in PINC. Hence, we focus on restricted 
versions of the maximal $\P$-subgraphs problem. For these restricted
versions, it is often easier to devise an efficient algorithm.
Later on we will show how, given an algorithm for one
of the restricted 
problems, the general problem can be solved. 

Let $G$ be a graph and let $\P$ be a property. 
We use $G-v$ to denote the induced graph of $G$ that contains all
vertices other than $v$. 
We say that  $G$ {\em almost satisfies\/} $\P$ if there is a
  vertex $v$ in $G$, such that 
  $G-v\in \P$. Let $v'$ be a vertex in $G$. 
We use $\maxSolVert \P G {v'}$ to denote the subset of
  $\maxSol \P G$ that contains graphs with the vertex $v'$. 

We will be interested in three restricted versions of the maximal
$\P$-subgraphs problem. 
\begin{itemize}
\item The {\em input-restricted maximal $\P$-subgraphs
    problem\/} is: Given a graph $G$ that almost
  satisfies $\P$, find all maximal $\P$-subgraphs of $G$.
\item The {\em output-restricted maximal $\P$-subgraphs
    problem\/} is: Given an arbitrary graph $G$ and a vertex $v'$ in $G$,
   find all maximal $\P$-subgraphs of $G$ that contain $v'$.
\item The {\em io-restricted maximal $\P$-subgraphs problem\/} is: Given
a graph $G$ that almost satisfies $\P$ and given a vertex $v'$ in $G$,
find all maximal $\P$-subgraphs of $G$ that contain $v'$.
\end{itemize}

Note that the output-restricted maximal $\P$-subgraphs problem can be
used in a straightforward way to solve the maximal $\P$-subgraphs
problem. However, it is not clear how the other two problems can be
used to solve the maximal $\P$-subgraphs problem. 

The complexity of these three problems is highly dependent on the
graph property $\P$. Sometimes, it turns out that the input-restricted
maximal $\P$-subgraphs problem and the io-restricted maximal
$\P$-subgraphs problem can actually be solved
in polynomial time, since the number of graphs in their output is
bounded in size by a constant. However, the 
fact that $G$ almost satisfies $\P$ does not always entail that
$\maxSol \P G$ is small. This is shown in the following example.

\eat{
The problems 

Interestingly, sometimes the 

For some properties $\P$ it turns out that if $G$ almost satisfies
$\P$, then the size of $\maxSol \P G$ is bounded by a constant.
However, sometimes the size of $\maxSol \P G$ (and even of
$\maxSolVert \P G {v'}$) may be
exponential in the size of $G$. We demonstrate with an example.
}

\begin{figure}[!t]
\begin{center}
\epsfig{figure=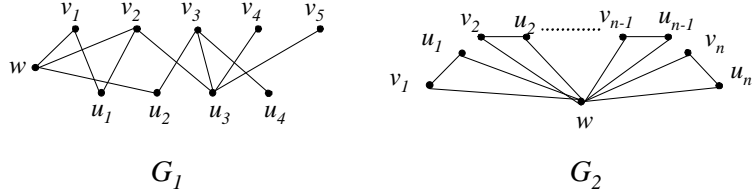,width=0.60\textwidth,clip=}
\end{center}
\caption{Finding Maximal Bipartite Subgraphs\label{fig:bipartite:graph}}
\end{figure}

\begin{example}
\label{example:pcbip:pbip}
Let $\PCBIP$ be the connected-hereditary property that contains all
connected bipartite graphs. Suppose that $G$ is a graph that almost
satisfies $\PCBIP$. It is not difficult to see that $\maxSol
{\PCBIP}G$ contains at most three graphs.
As an example, consider the graph $G_1$ in
Figure~\ref{fig:bipartite:graph}. The graph $G_1$ almost satisfies
$\PCBIP$, since $G_1\setminus w$ is a
connected bipartite graph. The set $\maxSol {\PCBIP}{G_1}$ contains
the following three graphs: {\em (1)\/} $G_1 \setminus w$, {\em (2)\/}
$G_1\setminus u_2$ (derived by removing the neighbors of $w$
on the bottom side) {\em and (3)\/} $G_1\setminus\set{v_1,v_2,u_1}$
(derived by removing the neighbors of $w$ on the top side,
and then removing unconnected vertices).

Let $\PBIP$ be the hereditary property that contains all
bipartite graphs. It is possible for the size of $\maxSol
{\PBIP} G$ to be  exponential in the size of $G$, even if $G$
almost satisfies $\PBIP$. Consider,
for example, the graph $G_2$ in Figure~\ref{fig:bipartite:graph}. The
graph $G_2$ almost satisfies $\PBIP$ since $G_2\setminus w\in
\PBIP$. However, the set $\maxSol \P {G_2} $ contains $2^n + 1$ graphs,
i.e., $G_2\setminus w$ and the graphs derived by choosing the vertex
$w$ and one from each pair of vertices $(v_i,u_i)$, for all
$i$. Notwithstanding the size of $\maxSol \P {G_2} $, it is not
difficult to show that the input-restricted $\PBIP$-subgraphs problem
is P-enumerable. To see this, observe that it is possible to find the
graphs in 
$\maxSol \P {G_2} $ efficiently, in terms of the input {\em and\/}
the output, by dealing separately with each connected component of
$G_2\setminus w$. 
\end{example}

By formalizing the intuition presented in Example~\ref{example:pcbip:pbip},
the following propositions can be shown. Similar propositions can be
shown for other graph properties.

\begin{proposition}
\label{proposition:restricted:pcbip}
The io-restricted $\PCBIP$-subgraphs problem is in \PTIME.
\end{proposition}

\begin{proposition}
\label{proposition:restricted:pbip}
The input-restricted $\PBIP$-subgraphs is {\rm P}-enumerable. 
\end{proposition}

\section{Hereditary Properties}\label{sec:hereditary}

In this section, we reduce the maximal $\P$-subgraphs problem to the
input-restricted maximal $\P$-subgraphs problem for hereditary properties
$\P$. Our reduction is by means 
of an algorithm that shows how to compute $\maxSol \P G$ for an
arbitrary graph $G$, given a procedure that can compute $\maxSol \P
G$ for graphs $G$ that almost satisfy $\P$. 

In Figure~\ref{fig:poly:alg:1}, the algorithm
{\sc GenHered}$\paramP$ is presented. 
This algorithm uses the following two procedures.
\begin{itemize}
\item {\sc Max}$\paramP(H,G)$: This procedure receives  graphs
  $H$ and $G$ 
  as input and returns true if $H$ is a maximal $\P$-subgraph of
  $G$ and false otherwise. This procedure can easily be defined in
  terms of {\sc 
    Sat}$\paramP$, by   {\em (1)\/} checking if $H\in \P$ {\em
    and (2)\/} extending $H$ with each vertex in
  $G$ and  checking whether any extension is in $\P$ (using
  {\sc Sat}$\paramP$).  
\item {\sc GenRestrHered}$\paramP(H)$: This procedure receives a graph $H$
  that almost satisfies $\P$ and returns the set $\maxSol \P H$. This
  procedure is not defined in this paper. Instead, it must be provided on a
per-property basis.
\end{itemize}
In essence, our algorithm  reduces the maximal
$\P$-subgraphs problem to the input-restricted maximal $\P$-subgraphs
problem by using the procedure {\sc GenRestrHered}$\paramP$. 

The algorithm  {\sc GenHered}$\paramP$ starts with the set $\G =
\set{\emptyGraph}$, where $\emptyGraph$ is an empty graph, i.e., a
graph with no vertices or edges. It then continuously (Line~6) attempts to
extend each graph $H$ in $\G$ with an additional vertex $v_i$. The
graphs in $\maxSol \P {G[H,v_i]}$ that are maximal with respect to the
vertices seen thus far are inserted into $\G$ (Line~7). 
This step is critical, since it {\em (1)\/} inserts graphs that are
needed in order to create the final result {\em and (2)\/} avoids
inserting extra graphs that would cause $\G$ to grow exponentially. 

Let $G$ be a graph with $n$ vertices. Suppose that there are $K$
graphs in $\maxSol \P G$. 
We show that {\sc GenHered}$\paramP(G)$
correctly computes $\maxSol \P G$ and  analyze the runtime of our
algorithm as a function of $n$ and $K$.
We use $s_\P(n)$ to denote the amount of time needed to check if 
$G\in\P$, 
i.e., the runtime of {\sc Sat}$\paramP(G)$.
Observe that {\sc Max}$\paramP(H,G)$ runs in $\O(n\,s_\P(n))$ time.
We use $r_\P(n,K)$ to denote the amount of  time needed to compute $\maxSol \P
G$, when $G$ almost satisfies $\P$, i.e., the runtime of {\sc
  GenRestrHered}$\paramP(G)$. Note that $r_\P$ is a function of both
the input and the output. 
\eat{
Finally, we use $\#_\P(n)$ to denote the maximum
number of graphs in $\maxSol \P G$, assuming that $G$ almost satisfies $\P$. 
Clearly, $\#_\P(n)$ is dominated by $r_\P(n,K)$, since $\#_\P(n)\leq
K$. However, we use the 
function $\#_\P(n)$ in our analysis since there are properties for
which $\#_\P(n)$ is a constant, e.g., $\PCBIP$. 
}

\begin{figure*}[t]
\begin{center}
\scalebox{0.9}{
\fbox{
\parbox{1in}{
\begin{tabbing}
0\=1\= 1. for\ \= for\ \= for\ \= for\ \= for\ \= AA \= \kill
\>\> {\bf Algorithm:} \> \>\>\> {\sc GenHered}$\paramP$ \ \ \ \\
\>\> {\bf Input:} \> \> \> \> 
Graph $G = (\set{\dd v n}, E, r)$ \\ 
\>\> {\bf Output:} \>\>\>\> Maximal $\P$-Subgraphs of $G$, i.e.,
$\maxSol \P G$  
\end{tabbing}\vspace{-0.5cm}
\begin{algorithm}{}{}
\G := \set{\emptyGraph} \\
\begin{FOR}{i := 1 \TO n}
  \H := \G \\
  \begin{FOREACH}{H\in \H}
      \G:= \G \setminus \set{H} \\ 
      \begin{FOREACH}{H'\in \CALL{GenRestrHered}\paramP(G[H,v_i])\ \ \ }
        \begin{IF}{\CALL{Max}\paramP(H',G[\set{\dd v i}])}
          \G := \G \cup \set{H'}
        \end{IF}
      \end{FOREACH}
  \end{FOREACH} 
\end{FOR} \\
\RETURN \G
\end{algorithm}
}
}
}
\end{center}
\caption{Algorithm to  compute $\maxSol \P {G}$ for
  hereditary properties
\label{fig:poly:alg:1}} 
\end{figure*}

%



\begin{theorem}
\label{theorem:chp:correct}
Let $\P$ be a hereditary property and let $G$ be a graph with $n$ vertices.
 Let $K$ be the number of graphs in $\maxSol \P G$. Then  
\begin{itemize}
\item {\sc GenHered}$\paramP(G) = \maxSol \P {G}$ {\em and\/}
\item {\sc GenHered}$\paramP(G)$ runs in time: 
$\O\big(n^2\,s_\P(n)\,K\,r_\P(n,K)\big)$. 
\end{itemize} 
\eat{{\sc GenHered}$\paramP(G) = \maxSol \P {G}$ and
{\sc GenHered}$\paramP(G)$ runs in time 
\[\O\big(n\, K\, (r_\P(n) + n\,s_\P(n)\,\#_\P(n))\big)\,.
 \]
}
\end{theorem}
\proofsketch
We use  $G_i$ to denote the induced subgraph of $G$ containing
exactly the vertices $\dd v i$, i.e., $G[\set{\dd v i}]$. We use
$\G_i$ to denote the value of $\G$ after $i$ iterations of the loop in
Line 2 of the algorithm. 
We show by
induction on the number of vertices $k$ in $G$ that 
$\G_k = \maxSol \P {G_k}$. The inclusion $\G_k \subseteq  \maxSol \P
{G_k}$ can be shown by a case analysis of the lines in which graphs
are added to (and removed from) $\G_k$. The inclusion $ \maxSol \P
{G_k} \subseteq \G_k$ follows from the following inclusion:
\begin{align*}
 \maxSol \P {G_{k}} \subseteq \maxSol \P {G_{k-1}} \cup \big(\cup_{H\in
                           \maxSol\P {G_{k-1}}}  \maxSol \P
                           {G[H,v_k]} \big)\,. 
\end{align*}
The runtime follows from a careful analysis of the algorithm and from
the fact that $|\maxSol \P {G_{i-1}}| \leq |\maxSol \P {G_i}|$, for
all $i$. 
\qed


\begin{corollary}
Let $\P$ be a hereditary property. 
 Then the maximal $\P$-subgraphs problem is in \PIO\  if and only if 
the input-restricted maximal $\P$-subgraphs problem is in \PIO.
\end{corollary}

\begin{corollary} \label{corollary:hereditary:P-enumerable}
Let $\P$ be a hereditary property. 
Then the maximal $\P$-subgraphs problem is {\rm P}-enumerable if
the input-restricted maximal $\P$-subgraphs problem is in \PTIME.
\end{corollary}

\begin{corollary}
The maximal $\PBIP$-subgraphs problem is in \PIO.
\end{corollary}

\renewcommand{\VCS}{{\sc GenWithVertex}$\paramP$}
\newcommand{\STACK}{{\mathit Stack}}

\section{Connected-Hereditary  and
  Rooted-Hereditary Properties}
\label{sec:connected-hereditary}
{\sc GenHered}$\paramP$ may fail to return the correct
graphs if $\P$ is  connected hereditary or
rooted hereditary. Intuitively, this 
failure is caused by the fact that an induced subgraph $H$ may not be
connected or rooted (and therefore, $H\not\in\P$), even though
there is a graph $G$ such that $H\gcon G$ and $G\in\P$.
In other words, the order in which we choose the vertices can
effect the success of the algorithm. 

In this section, we solve the maximal $\P$-subgraphs problem for
connected-hereditary and rooted-hereditary properties in the following
way:
\begin{itemize}
\item The maximal $\P$-subgraphs problem is reduced to the
  output-restricted maximal $\P$-subgraphs problem in a
  straightforward fashion.
\item The output-restricted maximal $\P$-subgraphs problem is reduced
  to the io-restricted maximal $\P$-subgraphs problem by means of the 
  algorithm \VCS\ (see Figure~\ref{fig:alg:vcs}). The result of calling
  \VCS$(G,v_r)$, for an arbitrary graph $G$, is the 
  set  $\maxSolVert \P G {v_r}$. Note that   \VCS$(G,v_r)$
  uses  {\sc
    GenRestrWithVertex}$\paramP(G,v_r)$  which generates
  $\maxSolVert \P {G'} {v_r}$ for graphs $G'$ that almost satisfy $\P$. 
\end{itemize}

In the remainder of this section, we explain the algorithm \VCS---its
 notation, data structures and flow of execution. We show its
 correctness and analyze its runtime.

\paragraph{Notation.}
Consider graphs $G$ and $H$ such that $H\gconeq
G$. 
We say that a vertex 
$v$ in $V(G)\setminus V(H)$ is an {\em undirected neighbor\/} of a
vertex $v'$ in $H$ if  either
 the edge $(v, v')$ or the edge $(v', v)$ is in $G$. Similarly, we say
 that $v\in V(G)\setminus V(H)$ is a {\em directed neighbor\/} of
 $v'$ in $H$ if 
 the edge $(v', v)$ is in $G$. 
Note that the neighbors (directed or undirected) of vertices in an
induced subgraph are {\em   not\/} in the induced subgraph. 

%
Given a property $\P$, 
we use $N_\P(H, G)$
to denote the set of 
undirected  neighbors of  $H$ if $\P$ is connected-hereditary
and to denote the set of directed neighbors of $H$ if $\P$ is
rooted-hereditary. Note that we use $\P$ only in order to differentiate between
undirected and directed neighbors. 

In our algorithm, a graph
 $H\gconeq G$ (such that $H\in \P$)
is associated with a set of vertices $\notV(H)$. Intuitively, this
set contains vertices $v'\in N_\P(H, G)$ that cannot be used to extend
$H$, since $G[H, v']\not\in \P$.

\paragraph{Data Structures.}
\VCS\ uses two stacks to collect graphs: $\STACK_1$ and
$\STACK_2$.
$\STACK_1$ contains graphs for which processing 
is incomplete. Therefore, a graph $H$ will be in $\STACK_1$ if it has a
 $v'\in N_\P(H, G)$ that is not in $\notV(H)$. For such a vertex $v'$, it is not
yet known whether $v'$ can be added to $H$, i.e., whether $G[H, v]\in
\P$.
$\STACK_2$ contains graphs for 
which processing is complete. Therefore, a graph $H$ will be in $\STACK_2$
if $N_\P(H, G)\subseteq \notV(H)$. 

To ensure that $\STACK_1$ and
$\STACK_2$ contain the proper graphs, our algorithm  uses the procedure
{\sc PushAppropriate}$\paramP(H,G, \STACK_1, \STACK_2)$,
  which does the following. If $N_\P(H, G) \not \subseteq \notV(H)$,
  then the procedure adds $H$ to the top of $\STACK_1$. Otherwise, the
  procedure adds $H$ to   the top of $\STACK_2$. 

\paragraph{Flow of Execution.}
The algorithm \VCS\ starts by considering the graph $G[\set{v_r}]$. Then, it
continually 
extends graphs in $\STACK_1$ with neighboring vertices to derive larger
graphs that are in $\P$. All 
extensions created must contain the vertex $v_r$ (so that we will only
create graphs in $\maxSolVert \P G {v_r}$). 
%
Suppose $H\in \STACK_1$ and $v\in N_\P(H, G)$.
We deal with the case in which
$G[H, v]\in \P$  in Lines~9-10.
We deal with the case in which  $G[H, v]\not\in \P$ in
Lines~11-25. \eat{Note that in the latter case we compute the  set
$\maxSolVert \P {G[H,v]} {v_r}$. One might think that 
 graphs $G'\in \maxSolVert \P
{G[H, v]} {v_r}$ can simply be added to $\STACK_1$ or $\STACK_2$, as
appropriate. However, this might cause an 
exponential blowup in the size of $\STACK_1$ and $\STACK_2$. Hence, we first
try to combine $G'$  with existing graphs in $\STACK_1$.
We also check if $G'$ is 
an induced subgraph of a graph in $\STACK_2$. Only if we have not
succeeded in either of these actions, 
$G'$ is added to $\STACK_1$ or $\STACK_2$, as appropriate. 
}

\paragraph{Correctness and Runtime Analysis.}
The proof of correctness of \VCS\ is rather intricate and has been
omitted due to lack of space. However, we take note of the behavior of our
algorithm that is critical in proving its correctness:
\begin{itemize}
\item The fact that $\notV(H)$ is assigned the empty set every time that
  a vertex is added to $H$, allows us to be prove that {\em all\/}
  graphs in $\maxSolVert \P G {v_r}$ are returned. 
\item Graphs $G'$  from $\maxSolVert \P {G[H,v]} {v_r}$ (see
  Line~12) are not immediately added to $\STACK_1$ or
  $\STACK_2$. Instead we first
try to combine $G'$  with existing graphs in $\STACK_1$.
We also check if $G'$ is 
an induced subgraph of a graph in $\STACK_2$. Only if we have not
succeeded in either of these actions, do we add
$G'$ to $\STACK_1$ or $\STACK_2$, as appropriate. This
prevents $\STACK_1$ and $\STACK_2$ from growing too big. It also ensures
that {\em only\/} graphs from $\maxSolVert \P G {v_r}$ are returned,
and each such graph is returned only once.
\end{itemize}
In order to prove our complexity analysis of the runtime of {\sc
  GenWithVertex}$\paramP$, we must show that {\sc
  GenRestrWithVertex}$\paramP$ does not 
 create more graphs than the number of graphs in
  the result of 
{\sc GenWithVertex}$\paramP$. This holds because we are able to prove
  that $H\gcon G$ implies
  that $|\maxSolVert \P   {G[H]} {v_r}| \leq 
|\maxSolVert \P {G} {v_r}|$   if either {\em (1)\/} $\P$ is connected-hereditary
{\em or (2)\/} $\P$ is rooted-hereditary and $v_r$ is the root of $G$.

\begin{figure*}[!t]\begin{center}
\scalebox{0.9}{
\fbox{\parbox{1in}{
\begin{tabbing}
0\=1\= AAA \= AAA \= AAA \= AAA \= AAA \= AA \= \kill
\>\>{\bf Algorithm:} \>\> \>\>{\sc GenWithVertex}$\paramP$\\ 
\>\>{\bf Input:} \> \> \>\> Graph $G$ and Vertex $v_r$ \\ 
\>\>{\bf Output:} \>\>\>\> Maximal answers that contain vertex $v_r$,
i.e., $\maxSolVert \P {G} {v_r}$ \ \ \ \ \ 
\end{tabbing}\vspace{-0.5cm}
\begin{algorithm}{}{}
\notV(G[\set{v_r}]) := \emptyset \\
 {\mathit Stack}_1 := \emptyset \\
 {\mathit Stack}_2 := \emptyset \\
 \CALL{PushAppropriate}\paramP(G[\set{v_r}], G, {\mathit Stack}_1, {\mathit Stack}_2) \\
\begin{WHILE}{{\mathit Stack}_1 \neq \emptyset}
  H:= {\mathit Stack}_1\CALL{.Pop()} \\
  \mbox{{\bf let} $v$ be a vertex in $N_\P(H, G)\setminus \notV(H)$} \\ 
  \begin{IF}{\CALL{Sat}\paramP(G[H, v])}
    \notV(G[H,v]):= \emptyset \\
    \CALL{PushAppropriate}\paramP(G[H,v]) 
    \ELSE \notV(H) := \notV(H) \cup\set{v} \\
    \begin{FOREACH}{G' \in \CALL{GenRestrWithVertex}\paramP(G[H,v],v_r)
    \setminus \set{H}} 
      \notV(G') := \emptyset\\
      {\mathit inserted} := \text{false} \\
      \begin{FOREACH}{H' \in {\mathit Stack}_1 \text{s.t.} \CALL{Sat}\paramP(G[G',H'])} 
        G_{\mathit new}:= G[G', H'] \\ 
        \notV(G_{\mathit new}):= \emptyset \\
        {\mathit Stack}_1.\CALL{Remove}(H') \\
        \CALL{PushAppropriate}\paramP(G_{\mathit new}, G, {\mathit Stack}_1, {\mathit Stack}_2) \\
        {\mathit inserted} := \text{true}
      \end{FOREACH}        \\
      \begin{IF}{\mbox{exists $H'\in {\mathit Stack}_2$ s.t. $V(G')\subseteq V(H')$}}
        {\mathit inserted} := \text{true}
      \end{IF}\\
      \begin{IF}{\mbox{not({\it inserted})}}
        \CALL{PushAppropriate}\paramP(G', G, {\mathit Stack}_1, {\mathit Stack}_2)
      \end{IF} 
    \end{FOREACH}\\
    \CALL{PushAppropriate}\paramP(H, G, {\mathit Stack}_1, {\mathit Stack}_2) 
  \end{IF} 
\end{WHILE}\\
\RETURN {\mathit Stack}_2
\end{algorithm}
}}}
\end{center}
\caption{Algorithm to compute $\maxSolVert \P G {v_r}$ for
  connected-hereditary or rooted-hereditary properties\label{fig:alg:vcs}}
\end{figure*}

Let $G$ be a graph with $n$ vertices.
We use $r'_\P(n,K)$ to denote the amount of time needed to compute
$\maxSolVert \P G v$ for a graph $G$ that almost satisfies $\P$ and an
arbitrary vertex $v$, i.e.,
the runtime of the procedure {\sc
  GenRestrWithVertex}$\paramP(G,v)$. Note that $r'_\P$ is a
function of both the input $n$ and the number of graphs in the output
$K$. The function $s_\P$ is defined as before.


\begin{theorem}
\label{theorem:vcs:correct}
Let $\P$ be a connected-hereditary or rooted-hereditary property. Let $G$
be a graph with $n$ vertices, and let $v_r$ be a vertex in
$G$.  Suppose that 
$G[\set{v_r}]\in\P$. Let $K$ be the number of graphs in $\maxSolVert \P G
v$.  Then 
\begin{itemize}
\item \VCS$(G, v_r) = \maxSolVert \P G {v_r}$ {\em and}
\item \VCS$(G, v_r)$ runs in time:
$\O\Big(n^2\,K^2\, r'_\P(n,K)\big(s_\P(n)\, +
n\big)\Big)$. 
\end{itemize}
\end{theorem}


\begin{theorem}
\label{corollary:vertex-polynomially-extendable:is:polynomially-solvable} 
Let $\P$ be a connected-hereditary or rooted-hereditary property.
The  maximal
  $\P$-subgraphs problem is in \PIO\ if the io-restricted maximal
$\P$-subgraphs problem is in \PIO.
\end{theorem}

\proofsketch
Since any vertex in $G$ may or may not appear in a 
solution of a connected-hereditary property, it is possible to compute
  $\maxSol \P G$ for connected-hereditary properties by calling \VCS\
  for every vertex in $G$. If $\P$ is a rooted-hereditary, every graph
  in $\maxSol \P G$ must contain the root of $G$. Hence, it is
  possible to compute $\maxSol \P G$ by calling \VCS\ with the root of
  $G$. 
\qed

If $\P$ is rooted
hereditary and $v_r$ is the root of
$G$, then $\maxSolVert \P G {v_r} = \maxSol \P
G$. Hence, we can show the following result.

\begin{corollary} \label{corollary:iff:rooted}
Let $\P$ be a rooted-hereditary property. The maximal
$\P$-subgraphs problem is in \PIO\ if and only if the 
input-restricted maximal $\P$-subgraphs problem is in \PIO. 
\end{corollary}

\section{Extending the Algorithms}

In this section we discuss some small changes that can be made to the
algorithms {\sc GenHered}$\paramP$ and \VCS\ in order to improve the
complexity results from the previous sections. 

\paragraph{P-Enumerable for Connected-Hereditary and Rooted-Hereditary
  Properties.} 
In Corollary~\ref{corollary:hereditary:P-enumerable}, we presented a
sufficient condition for the maximal $\P$-subgraphs problem to be
P-enumerable, for hereditary properties $\P$. The algorithm \VCS\
cannot be used in order to derive a sufficient condition for this
problem to be 
P-enumerable for connected-hereditary or rooted-hereditary
properties, since $K$ appears quadratically in the runtime of \VCS.

It turns  out that one can 
adapt  {\sc GenHered}$\paramP$ to derive an algorithm that computes
$\maxSolVert \P G {v_r}$, for a rooted-hereditary or
  connected-hereditary property $\P$,  provided that certain conditions hold.
The crux of the change to {\sc GenHered}$\paramP$ is in careful choice
of the order in which to iterate over the vertices in $G$. The adapted
algorithm can be used similarly to \VCS\ in order to compute $\maxSol \P G$. 

\begin{theorem}
Suppose that the input-restricted maximal
$\P$-subgraphs problem is in \PTIME. Then, the
maximal $\P$-subgraphs problem is {\rm P}-enumerable if
{\em (1)\/} $\P$ is rooted-hereditary and $G$ is acyclic
{\em or (2)\/} $\P$ is connected-hereditary and the underlying
undirected graph of $G$ is a tree.
\end{theorem}

\paragraph{Incremental Polynomial Time.}
None of the complexity results presented have provided conditions for
the maximal $\P$-subgraphs problem to be solvable in incremental
polynomial time. 
By slightly changing the procedure \VCS\
 we can  derive an algorithm that computes $\maxSolVert \P G
{v_r}$ in incremental 
polynomial time for an important special case. Using this adapted
algorithm,  $\maxSol \P G$ can also be computed in incremental 
polynomial time, for connected-hereditary and rooted-hereditary
 properties. 
Our adapted algorithm can also be used for a hereditary property $\P$, by
reducing $\P$ to an appropriately defined rooted-hereditary property. 
In addition, we derive a polynomial complexity result for returning
$k$ maximal induced subgraphs, for any constant $k$.

\begin{theorem}
\label{theorem:inc-poly}
Let $\P$ be hereditary, connected-hereditary or rooted-hereditary
property.  Suppose that the io-restricted maximal 
$\P$-subgraphs problem is in \PTIME. 
Then, 
{\em (1)\/} the maximal $\P$-subgraphs problem is in \PINC\ 
{\em and (2)\/} $k$ graphs from $\maxSol \P G$ can be returned in polynomial
  time, for any constant $k$.
\end{theorem}

\proofsketch 
Item~2 follows directly from Item~1. To show Item~1, 
let $G$ be a graph with $n$ vertices. 
By careful observation, one may note that after at most $n^2$
iterations of the loop in Line 5 of \VCS, an additional graph will
be in $\STACK_2$. One can take advantage of this fact to adapt \VCS\
so that it will run in incremental polynomial time, by having {\sc
  PushAppropriate}$\paramP$ print graphs as it adds them to
$\STACK_2$. (Care has to be taken not to print graphs that appeared
before in a previous call to \VCS.)
\qed

\begin{corollary}
The maximal $\PCBIP$ problem is in \PINC.
\end{corollary}

\eat{
This paper considers the problem of returning {\em all\/} maximal
solutions for a given property $\P$ and graph $G$. Another interesting
problem is that of returning $k$ maximal solutions for $\P$ and
$G$. Our incremental polynomial algorithm can be used to solve
 such problems in polynomial time, if the conditions of
 Theorem~\ref{theorem:inc-poly}  hold.
}
\eat{
\begin{corollary}
Let $G$ be a graph and $\P$ be a property. 
Suppose that either {\em
  (1)\/} $\P$ is hereditary or rooted-hereditary and the input-restricted
maximal $\P$-subgraphs problem is in \PTIME\  {\em or
  (2)\/} $\P$ is a 
  connected-hereditary property and the io-restricted maximal
  $\P$-subgraphs problem is in \PTIME.
 Then, it is possible to return $k$ graphs
from $\maxSol \P G$ in polynomial time, for any constant $k$.
\end{corollary}
}

\section{Conclusion} \label{sec:conclusion}

This paper reduces the maximal $\P$-subgraphs problem to restricted
versions of the same problem by providing algorithms that solve the
general problem, assuming that an algorithm for a 
restricted version is given. Our results imply that when attempting to
efficiently solve the maximal $\P$-subgraphs problem, it is not
necessary to define an algorithm that works for the general
case. Instead, an algorithm for restricted cases must be
defined.  An efficient  method for solving the maximal
$\P$-subgraphs problem for the general case is automatically
derived from our algorithms.

Sometimes it turns out that algorithms for restricted cases of the
maximal $\P$-subgraphs problem are straightfoward. For example, this
is the case with the properties $\PBIP$ and $\PCBIP$. There are
additional properties for which this holds, e.g., the set of
independent sets, the set of star graphs, etc. Thus, our results
immediately imply that the maximal \prop{is an independent
  set}-subgraphs problem is both P-enumerable and in \PINC,
and the maximal \prop{is a star
  graph}-subgraphs problem is in \PINC. Note
that it is significantly easier to come up with algorithms that solve
the restricted versions of these problems than to come up with 
algorithms that solve the general cases.

Interestingly, our results can be applied to 
the database problem of computing maximal query answers. Well-known
semantics for this problem, e.g., full
disjunctions~\cite{Galindo-Legaria:Full:Disjunctions}, can be modeled
as 
graph properties. It is often easy to define algorithms that solve the
restricted versions of the maximal $\P$-subgraph problem, for graph
properties that correspond to semantics for incomplete
information. Hence, the results in this paper have immediate practical
applications for efficiently computing maximal query answers.


\eat{Such restricted versions are often easier to
solve for specific properties $\P$ than the unrestricted version. Two
examples of properties for which an algorithm that solves a restricted
version is obvious are $\PBIP$ and $\PCBIP$.
For hereditary properties, the algorithm {\sc GenHered} reduces 
the maximal $\P$-subgraphs problem to the input-restricted maxmimal
$\P$-subgraphs problem. For connected-hereditary or rooted-hereditary
properties, the algorithm {\sc GenWithVertex} reduces the  maximal
$\P$-subgraphs problem to the io-restricted maxmimal $\P$-subgraphs
problem. These reductions give a complete characterization of when the
maximal $\P$-subgraphs problem is in \PIO, for hereditary and
rooted-hereditary properties, by reducing these problems to the
input-restricted maximal $\P$-subgraphs problem and the io-restricted
maximal $\P$-subgraphs problem, respectively.
If certain additional conditions hold, we have shown that the maximal
$\P$-subgraphs problem is in \PINC or is P-enumerable. 
}


\eat{
\begin{table}[t]
\begin{tabular}{lccc}
\toprule
Restricted in & Hereditary & Connected Hereditary & Rooted Hereditary
\\ \midrule 
input \PIO & \PIO & & \PIO \\  \midrule
io \PIO & \PIO & \PIO & \PIO \\  \midrule
input-restricted, 
 & \PINC, & \PINC, & \PINC, \\ 
io-restricted: \PTIME & P-enumerable &  P-enumerable  & P-enumerable  \\
& for arbitrary $G$  & if $G$ is acyclic & if $G$ is a tree \\
\bottomrule
\end{tabular}
\end{table}
}

\eat{
A complete
characterization is given for  when hereditary and
rooted-hereditary properties are in \PIO.

The goal of this paper is to investigate when all maximal solutions for
hereditary, connected-hereditary, and rooted-hereditary properties can
be computed in polynomial time under input-output complexity. 
Towards this end,  {\em
 polynomially-extendable\/} and {\em polynomially 
 vertex-extendible\/} properties are defined. The main
 results are 
 about hereditary properties that are also polynomially-extendable,
 and connected-hereditary or rooted-hereditary properties that are
 also polynomially vertex-extendible. In both of these
 cases, all maximal solutions can be generated in polynomial time
 under input-output complexity.  We show that all
 polynomially-solvable hereditary properties are
 polynomially-extendable and all polynomially-solvable
 rooted-hereditary properties are polynomially
 vertex-extendible. It is often easy to verify whether a
 property is polynomially extendible or polynomially
 vertex-extendible. In such cases, one can immediately determine whether a
 property is polynomially solvable. 

For hereditary and rooted-hereditary properties $\P$, such that
$\funExtend n m$ is polynomial in $n$ and $m$, and for
connected-hereditary properties $\P$, such that $\funExtendVert n m$ is
polynomial in $n$ and $m$, we present an incremental polynomial
algorithm that generates all maximal solutions. For such properties,
we also show that $k$ maximal solutions can be generated in polynomial
time in the size of the input, for a constant $k$. 

We now present examples of hereditary, connected-hereditary and
rooted-hereditary properties which were studied independently in the
past. For these properties, we demonstrate how our results contain and
improve upon previous results. 

\paragraph{Graph Theory.}
One application where it is of importance to find all maximal solutions
to a hereditary property is that of graph coloring.
There are  algorithms for coloring a graph that 
first compute all maximal independent sets of the graph or all maximal 
bipartite subgraphs of the graph. An algorithm for computing maximal
independent sets was presented
in~\cite{Johnson-Yannakakis:Generating-Maximal-Independents-Sets}.
In~\cite{Number:Bipartite:Graphs}, an
algorithm for finding all maximal bipartite subgraphs of a graph is
presented. This algorithm computes all maximal bipartite subgraphs of
a graph $G$ with $n$ vertices in time proportional to the maximal
number of bipartite subgraphs that a graph of size $n$ may
have. However, this algorithm is not guaranteed to be polynomial in
the size of the actual input and output. Since \prop{is a bipartite
  graph} is a polynomially-extendable hereditary property, our
algorithm \CHP, presented 
in Section~\ref{sec:hereditary},
computes all maximal bipartite subgraphs in
polynomial time under input-output complexity.

\eat{
\paragraph{Classical Database Theory.}
Many database problems  involve
hereditary properties. For example, in order to  check if a
relation $R$ is in BCNF, one must generate
all minimal keys of $R$. The characteristic of being a minimal key
is closely related to a hereditary property, since the property
\prop{is a  complement of a minimal key} is hereditary. 
Many data mining tasks, such as finding frequent
itemsets, involve
generating maximal sets with a given property. 
Although the problems discussed in this subsection involve hereditary
properties, it is not yet clear if it is possible to apply our techniques
to these problems.
}

\paragraph{Computing Maximal Answers to Queries.}
With the development of the World-Wide Web and the high availability
of data from widely varying sources, the problem of integrating
information from heterogeneous sources has received much attention. In
such scenarios, it is common for the information available to be
partial. In general it may be the case that there is not enough
information to completely answer (satisfy) a query. Under such
conditions, one is interested in finding all maximal answers for a
given query.  

Full disjunctions are a well-known  method used to answer queries in incomplete
relational databases
(see~\cite{Rajaraman:Ullman-Computing:Full:Disjunctions-PODS,Galindo-Legaria:Full:Disjunctions}). 
Only recently, \cite{Kanza:Sagiv:Full:Disjunctions} showed that 
 full disjunctions can be computed in polynomial time under input-output
complexity. 
We show in Appendix~\ref{appendix:examples} that  a full disjunction
can  be computed 
by generating maximal solutions for a suitably defined
connected-hereditary property.  A precise definition of a full disjunction
and of the corresponding
connected-hereditary property $\P_{\sf fd}$
 is presented in Appendix~\ref{appendix:examples}. The property
 $\P_{\sf fd}$ is polynomially vertex-extendible and $e'_{\P_{\sf
     fd}}(n, m)$ is polynomial in $n$ and $m$. 
Therefore, this
paper improves on the result in~\cite{Kanza:Sagiv:Full:Disjunctions}
since the algorithm in Section~\ref{sec:ip:connected-hereditary} can
be used to compute $\P_{\sf fd}$ in incremental polynomial  time.

In~\cite{Kanza:Et:Al-Incomplete:Answers:Over:SSDs-PODS} the problem of
computing maximal answers for rooted graph queries over
rooted graph databases was considered. For two of the semantics
presented, {\em {\sc or}-semantics} and {\em weak semantics},
algorithms that generate all maximal answers in  polynomial time
under input-output complexity were
presented~\cite{Kanza:Et:Al-Incomplete:Answers:Over:SSDs-PODS,Kanza:Sagiv:Full:Disjunctions}.
One can compute maximal
answers under weak semantics and under
{\sc or}-semantics  by computing maximal solutions for
appropriately defined rooted-hereditary properties. The definition of
weak semantics and {\sc or}-semantics, along with 
rooted-hereditary properties $\P_{\sf weak}$ and $\P_{\sf or}$ for
these problems, are presented in Appendix~\ref{appendix:examples}. 
The algorithms
presented in this paper for generating maximal solutions for
rooted-hereditary properties can be used to solve these problems. 
Since {\em (1)\/} $\P_{\sf weak}$ and $\P_{\sf or}$ are polynomially
vertex-extendible {\em and (2)\/} $e_{{}_{\P_{\sf weak}}}(n,m)$ and
$e_{{}_{\P_{\sf or}}}(n,m)$ are polynomial in $n$ and $m$, it is possible
to compute all maximal answers under weak semantics and under {\sc
  or}-semantics in incremental polynomial time, which is an
improvement on previous results.

The problem of extracting
maximal tuples (i.e., tuples padded with null values) from XML
documents was considered in~\cite{Cohen:Generating:Relation}. They
generated maximal tuples that are {\em completely interconnected\/} or
{\em reachably interconnected} in polynomial time under input-output
complexity. It is fairly straightforward to define a hereditary
property that corresponds to complete interconnectedness and a
connected-hereditary property that corresponds to reachable
interconnectedness. Hence, our algorithms can be used to generate
their maximal tuples. In addition, our algorithms are superior to
those in~\cite{Cohen:Generating:Relation} in their
runtime. Furthermore, our incremental polynomial algorithms can be
used to generate maximal tuples in incremental polynomial time. 

\enlargethispage{\baselineskip}
In summary, we believe that this paper lays a theoretical foundation for the
problem of defining query semantics over incomplete information. The
results in this paper imply that when  defining a semantics for
incomplete query answers it is often 
useful to define the semantics in terms of a hereditary,
connected-hereditary or rooted-hereditary property that is 
polynomially solvable. In fact, we have shown that previously defined
semantics for incomplete information were  unwittingly defined in
this way. We leave for future work the task of  defining a semantics
for incomplete 
information for XQuery, based on the principles presented in this paper.

\eat{
One open question is whether is is possible for a property to be
connected-hereditary, polynomially solvable and {\em not\/} polynomially
vertex-extendible. 
}
}

\bibliography{strings,%
              hereditary-lit}
\bibliographystyle{abbrv}

\end{document}